\documentclass{pasj00}

\newcommand{\lsim}{\raisebox{0.3mm}{\em $\, <$} \hspace{-3.3mm}
\raisebox{-1.8mm}{\em $\sim \,$}}
\newcommand{\gsim}{\raisebox{0.3mm}{\em $\, >$} \hspace{-3.3mm}
\raisebox{-1.8mm}{\em $\sim \,$}}

\begin{document}
\SetRunningHead{Ohsuga, Susa, \& Uchiyama}{Population III Disks}
\Received{2007/04/09}
\Accepted{2007/07/27}

\title{Instability of Population III Black Hole Accretion Disks}

\author{Ken \textsc{Ohsuga},\altaffilmark{1,2}
        Hajime \textsc{Susa},\altaffilmark{2,3}
        and
        Yosuke \textsc{Uchiyama}\altaffilmark{2}
        }
\altaffiltext{1}{Institute of Physical and Chemical Research (RIKEN), 
2-1, Hirosawa, Wako, Staitama 351-0198}
\altaffiltext{2}{Department of Physics, Rikkyo University, 
Toshima-ku, Tokyo 171-8501}
\altaffiltext{3}{Department of Physics, Konan University
8-9-1 Okamoto, Kobe 658-8501}

%
%


\maketitle


\begin{abstract}
We investigate the stability of black hole accretion disks 
in a primordial environment (POP III disks for short),
by solving the vertical structure of optically thick disks,
including convective energy transport,
and 
by employing a one-zone model for optically thin isothermal disks.
Because of the absence of metals in POP III disks, 
we find significant differences in stability associated with ionization 
between POP III disks and the disks of solar metallicity.
An unstable branch in S-shaped equilibrium curves on the $\dot{M}-\Sigma$ 
(mass accretion rate - surface density) plane
extends to a larger surface density compared with the case of 
disks of solar metallicity. 
The resulting equilibrium loci indicate that 
quasi-periodic oscillations in luminosity can also be driven 
in POP III disks, and their maximal luminosity 
is typically by an order of magnitude larger than 
that of the disks of solar metallicity. 
Such a strong outburst of POP III disks
can be observed by future huge telescopes,
in case that the mass is supplied onto the disks at the Bondi accretion
rates in typical virialized small dark halos.
\end{abstract}



\section{Introduction}
According to recent advances in the formation theory of population
III (POP III) stars, it has been understood that they could be 
very massive $(10^2-10^3M_{\odot })$ \citep{ON98,Abel00,BCL02}.

Those stars end up as black holes as massive as $10-10^3 M_\odot$ 
(here after POP III BHs), 
in the case that the mass of the progenitor stars is within the range
$40M_\odot \lsim M_* \lsim 140M_\odot$ or $M_*\gsim 260M_\odot$ 
 \citep{HW02}.
The accretion disks formed around such black holes can emit a significant 
amount of soft X-rays, which can contribute to reionization of the
universe \citep{Ricotti04}. Moreover, such black holes could be 
candidates for the progenitors of ultra-luminous X-ray
sources \citep{Mii05} 
or the building blocks of super-massive black holes \citep{Rees84}. 

The accretion disks surrounding POP III BHs are formed from the accreting gas
around the progenitor stars.  In the case that the mass of a POP III
star is within the range 
$40M_\odot \lsim M_* \lsim 140M_\odot$ or $ M_* \gsim 260M_\odot$, 
the star forms a black hole directly without a SN \citep{HW02}. 
In this case, the accreting gas is expected to contain 
a small amount of heavy elements. 
In addition, recent numerical 
simulations on low-mass
galaxy formation, taking into account the effects of SNe feedback,
suggest that the metallicity of the gas remain in such galaxies during
their early star-burst phase is as low as $\sim 10^{-4} Z_\odot$
\citep{wada03}. 
Therefore, the metallicity of the accretion disks
surrounding POP III BHs is expected to be very low 
($Z \lsim 10^{-4}Z_\odot$).
We hereafter call these accretion disks as POP III disks.
 
The thermal instability of the standard accretion disk has been
investigated in detail 
(e.g, \authorcite{SH75} \yearcite{SH75}; 
\authorcite{SS76} \yearcite{SS76}; 
\authorcite{Pringle76} \yearcite{Pringle76}).
By assuming solar metallicity (hereafter POP I disks),
Mineshige and Osaki (\yearcite{MO83})
obtained the S-shaped equilibrium loci on 
the $\dot{M}-\Sigma$
(mass accretion rate v.s. surface density) plane,
implying that 
the unstable disk undergoes limit-cycle oscillation
between the upper, hot, ionised branch and the lower, cool
neutral branch 
(see also 
\authorcite{Hoshi79} \yearcite{Hoshi79};
\authorcite{Meyer81} \yearcite{Meyer81};
\authorcite{Smak82} \yearcite{Smak82};
\authorcite{aobon} \yearcite{aobon} for a review).
It is so-called dwarf-nova type instability,
which successfully explains the observed properties of dwarf nova
\citep{Meyer84}.
An X-ray nova is also thought to be caused by 
such a disk instability, although its central object is
expected to be a black hole \citep{HW89,MW89}.
The limit-cycle behavior for active galactic nuclei
has been investigated by \citet{LS86} 
(see also \authorcite{MS90} \yearcite{MS90}).

Mayer and Duschl (\yearcite{MD05a})
investigated this instability on the POP
III disks in which they also obtained the equilibrium loci 
on the $\dot{M}-\Sigma$ plane. 
However, their results based upon a one-zone
approximation,
even if the disk is optically thick,
in which the energy transfer due to 
convection along the direction perpendicular to the equatorial plane of the
disk is not included. 
As a result, the equilibrium loci are not S-shaped, which does not directly
lead to the limit-cycle oscillations driven by thermal and secular
instabilities. 
Thus, we need a more sophisticated
approach in order to understand the nature of the stability of POP III disks.
In this paper, we re-examine stability of POP III disks,
by solving the vertical structure of optically thick disks, 
including the effect of convection.
For optically thin disks, 
we adopt a one-zone approximation.
In addition, we evaluate
the amplification of the luminosity of accretion disks surrounding
POP III BHs by the limit-cycle oscillation.
We also discuss the possibility to detect the POP III disks
by calculating their emergent spectra in the burst phase.
 
This paper is organised as follows:
we describe basic equations in section 2. 
We compare the stability of POP III disks with POP I disks in section 3.
In section 4, we discuss the observational possibility of POP III black
hole accretion disks, and summarise in the final section.

\section{Model and Numerical Method}
\subsection{Optically Thick Regime}
\label{thick}
In this subsection, 
we show the basic equations used to describe 
the vertical structure of the optically-thick accretion disks.
Here, we assume that the disks are in hydrostatic equilibrium along
the vertical axis, and local energy balance is attained.
The convective energy transport is taken into consideration.
The equation of hydrostatic equilibrium is given by
\begin{equation}
 \frac{dp}{dz}=-\rho\frac{GMz}{r^3},\label{eq:fbalance}
  \label{hydro}
\end{equation}
where $p$ is the total pressure,
$\rho$ is the density, 
$M$ is the black hole mass,
and $r$ and $z$ are the radial and vertical coordinates.
The equation of energy transport is
\begin{eqnarray}
 \frac{d\ln T}{d\ln z}=
  \left\{ 
   \begin{array}{ll}
    \nabla_{\rm rad} & \left(\nabla_{\rm rad} < \nabla_{\rm ad}\right) \\
    \nabla_{\rm conv} & \left(\nabla_{\rm rad} > \nabla_{\rm ad}\right) \\
   \end{array}
  \right.,
  \label{temp}
\end{eqnarray}
where $T$ is the temperature, and
$\nabla_{\rm ad}$ is the adiabatic gradient. 

The radiative gradient, $\nabla_{\rm rad}$, is written as
\begin{equation}
 \nabla_{\rm rad}=\frac{3r^3\kappa p F}{4acT^4GMz},
\end{equation}
where $\kappa$ represents the opacity,
$F$ and $a$ denote the energy flux and the radiation constant, 
respectively.
We utilise the latest OPAL opacity 
table\footnote{http://www-phys.llnl.gov/Research/OPAL/opal.html} including
low temperature opacities \citep{OPAL05}.
The energy flux is evaluated by integrating the following equation, 
which represents the local energy balance :
\begin{equation}
 \frac{dF}{dz}=-\frac{3}{2}\alpha p \Omega.
  \label{flux}
\end{equation}
Here $\alpha$ is the viscosity parameter, whereas 
$\Omega=(GM/r^3)^{1/2}$ represents the Keplerian angular speed.
We employ the expression 
$\alpha=\min[1.0, 10^2(H/r)^{1.5}]$
(with $H$ being the disk half thickness), 
which succeeds to reproduce the variation 
amplitude and the timescale 
of the X-ray novae 
(\authorcite{aobon} \yearcite{aobon}, see also
\authorcite{Meyer84} \yearcite{Meyer84}),
whereas the viscosity model is still poorly understood for
the primordial composition.
We here note that the source of disk viscosity 
is though to be of magnetic origin based on the recent 
magnetohydrodynamic simulations
\citep[for a review]{HBS01, MMM01, Balbus03}.

The convective gradient, $\nabla_{\rm conv}$, is 
calculated by the mixing-length formalism for
convection  \citep{Pac69}, in which we assume 
mixing length, $l$, to be
\begin{equation}
 l = \min(H,H_{\rm p}),\label{eq:mix}
\end{equation}
where $H_{\rm p}$ is the pressure scale height.

In order to close the set of equations (\ref{eq:fbalance})-(\ref{eq:mix}),
we need the equation of state for the system in radiative equilibrium,
\begin{equation}
 p=\frac{\rho kT}{\mu m_{\rm p}}+\frac{1}{3}aT^4,
\end{equation}
where $m_{\rm p}$ denotes the proton mass,
$k$ represents the Boltzmann constant,
and $\mu$ is the mean molecular weight.

We integrate equations (\ref{hydro}), (\ref{temp}), and (\ref{flux})
from the disk surface ($z=H$) toward the equatorial plane ($z=0$)
by the Runge-Kutta method for given parameters,
$r$, $Z$, and $\dot{M}$, 
where $Z$ is the metallicity and
$\dot{M}$ is the mass accretion rate.
The outer boundary ($z=H$) conditions for temperature,
energy flux, pressure are given by
\begin{equation}
 T = \left(\frac{3}{8\pi}\frac{GM\dot{M}}{\sigma r^3}\right)^{1/4},
  \label{Teff}
\end{equation}
\begin{equation}
 F=\sigma T^4,
\end{equation}
and
\begin{equation}
p = \frac{2}{3\kappa}\Omega^2 H,
\end{equation}
where $\sigma$ denotes the Stefan-Boltzmann constant.
We iteratively search $H$
so as to meet the condition $F=0$ at $z=0$.
Then, we obtain the vertical structure of the accretion disks.
In this method, the column density and the optical thickness
are calculated from $\Sigma=2 \int_0^H \rho dz$
and $\tau=\int_0^H \rho \kappa dz$,
respectively.
This method breaks down 
in the regions of a small mass accretion rate
for POP III disks, since then the disks become
optically thin (We will discuss this point later).
A numerical method for optically thin disks is 
described in the next subsection.

\subsection{Optically Thin Regime}
\label{thin}
The optically thin disks would be nearly isothermal,
in which convection does not contribute to the energy 
transport and the radiative diffusion approximation 
breaks down.
Here, we adopt a one-zone model for optically thin disks.
Since the radiation pressure is negligible in the optically thin 
medium, the equation of hydrostatic equilibrium is given by
\begin{equation}
 \Omega^2 H^2 = \frac{p_{\rm gas}}{\rho},
  \label{1Dhydro}
\end{equation}
where $p_{\rm gas}=\rho kT/\mu m_{\rm p}$ 
is the gas pressure.
The viscous heating rate is described as 
\begin{equation}
 Q_{\rm vis}=
  -\frac{3}{2}T_{r\varphi}\Omega,
  \label{Qvis}
\end{equation}
where $T_{r\varphi}$ is the shear stress tensor.
Using the alpha-prescription, 
$T_{\rm r\varphi}=-2\alpha p_{\rm gas} H$,
we rewrite the equation (\ref{Qvis}) as 
\begin{equation}
 Q_{\rm vis}=
  \frac{3}{2}\alpha\frac{p_{\rm gas}}{\rho}\Omega\Sigma,
  \label{Qvis2}
\end{equation}
where $\Sigma=2\rho H$ is the column density.
The viscosity parameter, $\alpha$, is given by
$\alpha=\min[1.0, 10^2(H/r)^{1.5}]$,
the same as in the method for the optically thick disks.
The radiative cooling rate is written as
\begin{equation}
  Q_{\rm rad}= 2\sigma T^4\tau,
  \label{Qrad}
\end{equation}
where $\tau=\kappa\Sigma/2$ is 
the optical thickness of the disk.
Combining the energy balance equation ($Q_{\rm vis}=Q_{\rm rad}$)
and equation (\ref{1Dhydro}), 
we obtain the physical quantities of the disks.

To calculate the mass-accretion rate, we use 
the continuity equation 
($\dot{M}={\rm const.}$ in $r$-direction)
and the angular-momentum conservation low.
By assuming a torque-free boundary condition
and angular momentum of the gas to be much smaller 
at the inner boundary than at the outer region,
we have 
\begin{equation}
 T_{r\varphi}=-\frac{1}{2\pi}\sqrt{\frac{GM}{r^3}}\dot{M}.
  \label{T2}
\end{equation}
Thus, using equation (\ref{Qvis}) and (\ref{T2}),
the mass-accretion rate is give by
\begin{equation}
 \dot{M}=\frac{4\pi}{3}\frac{r^3}{GM}
 Q_{\rm vis}.
 \label{1Dmdot}
\end{equation}

Throughout the present study, 
we set the black hole mass to be $M=10^3M_\odot$,
unless stated otherwise.

\section{Results}
\label{sequence}
\begin{figure}
  \begin{center}
    \FigureFile(90mm,90mm){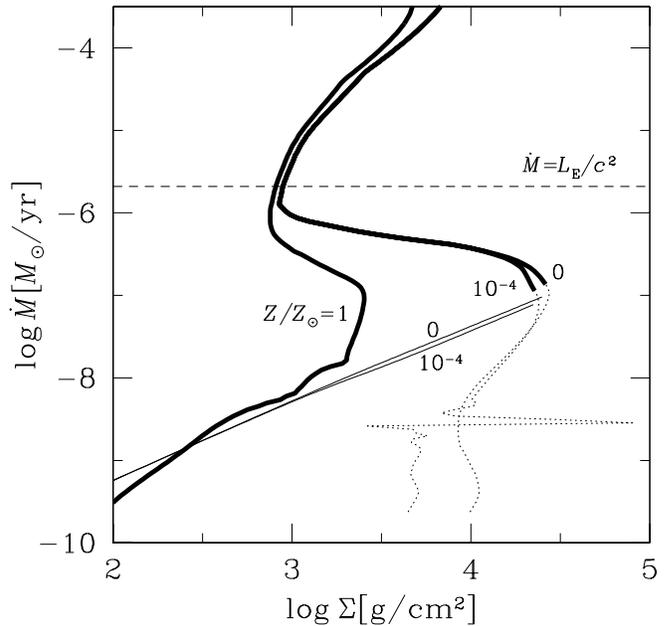}
  \end{center}
 \caption{
 Sequence of equilibrium curves of the optically-thick accretion disks
 with $\tau \geq 3$ ({\it the thick solid lines}) 
 and the optically-thin isothermal disks with $\tau\leq 0.3$
 ({\it the thin solid lines})
 for $Z/Z_\odot=0$, $10^{-4}$, and $1$.
 The dotted lines are equilibrium loci for POP III disks 
 with $\tau<3$, which is obtained by the method described 
 in \S\ref{thick}, 
 although this method might break down there.
 The dashed line indicates the critical mass-accretion rate,
 $L_{\rm E}/c^2$, where $L_{\rm E}$ is the Eddington luminosity.
 The adopted parameters are 
 $M=10^3M_\odot$ and $r=10^4 r_{\rm S}$.}
 \label{Sji}
\end{figure}

In figure \ref{Sji}, we present the sequence 
of equilibrium curves 
on the $\dot{M}-\Sigma$ 
(mass accretion rate - surface density)
plane for $r=10^4r_{\rm S}$ with 
$Z/Z_\odot=0$, $10^{-4}$, and $1$,
where $r_{\rm S} \equiv 2GM/c^2$ denotes the Schwarzschild radius.
Here, the thick and thin solid lines indicate the resulting 
equilibrium loci for optically thick accretion disks
with $\tau\geq 3$,
in which the energy is transported via convection 
as well as radiative diffusion (see \S\ref{thick}),
and the optically thin isothermal disks with $\tau\leq 0.3$
(see \S\ref{thin}).
Our numerical method for optically thick disks 
gives optically thin branches for POP III disks
when $\dot{M} \lsim 10^{-7}M_\odot/{\rm yr}$.
Although they are unphysical branches,
we represent them by the dotted lines.

We find that 
the optically-thick POP III disks transit 
to the optically-thin isothermal disks 
at around $\dot{M}\sim 10^{-7}M_\odot {\rm yr}^{-1}$.
In contrast, 
the POP I disks stay
optically thick ($\tau\geq 3$),
even if the mass-accretion rate is less than
$10^{-7}M_\odot {\rm yr}^{-1}$.
We find that all curves have two major turning points
at $\dot{M}\sim 10^{-7} M_\odot{\rm yr}^{-1}$
(lower-right turning point) and 
$10^{-6} M_\odot{\rm yr}^{-1}$
(upper-left turning point),
where the gradients of the curves, 
$\partial \dot{M}/\partial \Sigma$, change their signs.
In other words, the S-shaped equilibrium curves appear 
irrespective of the metallicity.
It is known that the middle branch ($\partial \dot{M}/\partial \Sigma <
0$~:~ between the upper-left and lower-right turning points)
is thermally and secularly unstable,
whereas the disk is stabilised in upper 
and lower branches, along which $\partial \dot{M}/\partial \Sigma > 0$
being satisfied.

Figure \ref{Sji} indicates that 
if the mass-accretion rate falls 
within the range $10^{-7} M_\odot{\rm yr}^{-1}
\lsim \dot{M} \lsim 10^{-6} M_\odot {\rm yr}^{-1}$,
the POP III disks exhibit the limit-cycle behaviour, 
as well as the POP I disks \citep{MO83},
since no stable branch exists.
Note that the range of the mass-accretion rate, 
which induces the limit-cycle behavior, 
increases as the radius increases.
We discuss this point in \S 4.1.
When the disk stays in the lower stable branch,
the surface density at $r=10^4r_{\rm S}$ increases with time, because 
the mass input rate 
(mass supplied from outside per unit time)
is larger than the mass-output rate 
(mass ejected inward per unit time).
As a result, the disk evolves along the lower stable branch
toward the lower-right turning point
on the $\dot{M}-\Sigma$ plane (quiescent state).
After a while, it suddenly jumps to the upper branch at the right end of 
the stable branch (lower-right turning point).
Since the mass-input rate is smaller than 
the mass-output rate in the upper stable branch, 
$\Sigma$ decreases along the upper stable branch
toward the upper-left turning point (burst state)
until it jumps back to the lower branch.
Thus, quasi-periodic oscillation of 
the mass-accretion rate onto the black hole
is caused by such state transitions of the disk,
leading to oscillations in luminosity.

Here, we stress that the S-shaped equilibrium curve 
of the POP III disks, as well as the POP I disks,
does not appear if we employ the one-zone approach
for optically thick disks 
(see figure 4 in \authorcite{MD05a} \yearcite{MD05a}).
It is crucial to solve the vertical structure of 
the optically thick disks, including the convective effect,
in order to investigate the bursting phenomenon 
caused by a dwarf-nova type instability.

Although the limit-cycle oscillations could arise 
both in POP I and POP III disks,
the POP III disks exhibit stronger outbursts than do the POP I disks.
The surface density at the lower-right turning point
is about an order of magnitude 
larger in the POP III disks than in the POP I disks,
whereas the upper stable branch is independent of the metallicity.
Consequently, the luminosity of POP III disks is 
about an order of magnitude larger than
that of the POP I disks just after a state transition from the quiescent 
to burst states.
As shown in figure \ref{Sji},
the equilibrium loci for $Z/Z_\odot=0$ and $10^{-4}$ 
almost overlap each other.
It implies that the magnitude of the outburst 
is independent of the metallicity
as long as $Z/Z_\odot\lsim 10^{-4}$ is satisfied.
Here, we note that 
the S-shaped equilibrium curves 
for $Z/Z_\odot \gsim 10^{-3}$
deviate significantly from those for the POP III disks, 
although we do not plot them in this figure. 
Whereas the upper branch does not depend upon $Z$ very much,
the lower turning point shifts toward the left side
as the metallicity increases for $Z/Z_\odot \gsim 10^{-3}$.

Since we have $\alpha \sim 1$ and $T \sim 10^5 \rm K$
in the upper branch, 
the timescale of the burst phase is evaluated to be 
\begin{eqnarray}
 t_{\rm b}\sim 2.1 \alpha^{-1} & &
  \left(\frac{M}{10^3M_\odot}\right)^{1/2}\nonumber\\
  & &\times\left(\frac{r}{10^4r_{\rm S}}\right)^{1/2}
  \left(\frac{T}{10^{5}{\rm K}}\right)^{-1}{\rm yr}.
\end{eqnarray}
This timescales is basically the viscous timescale.
The timescale for the quiescent phase
is given by $\pi r^2 \Sigma/\dot{M}$.
It is described as 
\begin{eqnarray}
 t_{\rm q}\sim  & & 4.2\times 10^2
  \left(\frac{r}{3\times 10^{12}{\rm cm}}\right)^2 \nonumber\\
  & &\times \left(\frac{\Sigma}{3\times 10^{4}{\rm g\,cm^{-2}}}\right)
  \left(\frac{\dot{M}}{10^{-6}M_\odot\,{\rm yr}^{-1}}\right)^{-1}
  {\rm yr}.
\end{eqnarray}
The disc evolves along the lower stable branch on this timescale, and it
attains the lower-right turning point.
The timescales of the burst and quiescent phases 
are much longer than those of the transition 
between the burst and quiescent states.
Hence, about 
$\epsilon=t_{\rm b}/t_{\rm q}\sim 0.5\%$ 
of the unstable POP III disks
would stay in the burst state, which are observed as 
very luminous objects.
Here, we remark that global calculations of the disks are
need for investigating the time evolution of the disks in detail.

In figure \ref{Sji}, we find that the mass-accretion rate
exceeds the critical rate in the burst phase.
In this phase, $H/r$ is comparable to, or slightly larger 
than, unity, although the disk is assumed to be geometrically 
thin in our method.
The radiation pressure-dominated region
would appear at the inner region of the disk,
in which the disk is unstable 
if the viscosity is proportional to the total pressure.
Such coupling of two instabilities has been suggested
by Lin and Shields (\yearcite{LS86}).
By supercritical accretion,
the radiatively driven outflows would form 
in the vicinity of the black hole,
and the photon would be trapped in the flow within 
the trapping radius, $r_{\rm trap}\sim (\dot{M}c^2/L_{\rm E})r_S$
\citep{Ohsuga02}.
We need a global multi-dimensional approach to
investigate the disk structure in detail
(e.g., \authorcite{Ohsuga05} \yearcite{Ohsuga05};
\authorcite{Ohsuga06} \yearcite{Ohsuga06}; \yearcite{Ohsuga07}).
However, it is beyond the scope of the present study.

The irradiation flux
is not taken into consideration in the present study.
\citet{TMW90} has reported that 
convective disks are stabilized by strong 
irradiation with $T>10^4{\rm K}$.
However, it is not easy to get the irradiation flux,
since it is very sensitive to the geometry and the radiative processes.
In the quiescent phase, the disk is very geometrically thin,
so that the irradiation flux would be reduced.
In the burst phase, the thick disk forms via the supercritical accretion.
Then, the radiative flux is collimated in the polar direction,
suppressing the irradiation effect \citep{Ohsuga05}.
The self-occultation by the thick disk
would also reduce the irradiation flux \citep{Watarai05}.
For this issue, we should study the global structure of the disks, 
fully accounting for radiative transfer in the multi-dimensional space.

\begin{figure}
  \begin{center}
    \FigureFile(90mm,90mm){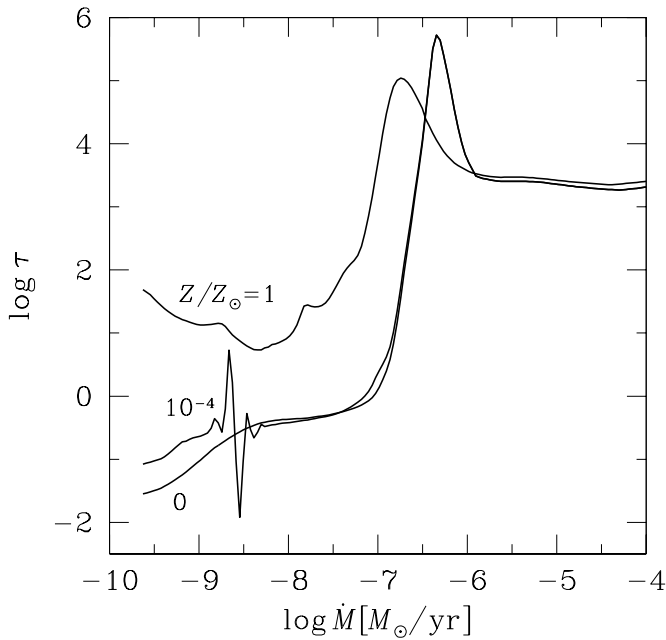}
  \end{center}
 \caption{
 Optical thickness of disks
 for $Z/Z_\odot=0$, $10^{-4}$, and $1$
 as functions of the mass-accretion rate,
 where we use the method described in \S\ref{thick}. 
 Our method, in which we assume the disk to be optically thick,
 breaks down for POP III disks with 
 $\dot{M}\lsim 10^{-7}M_\odot {\rm yr}^{-1}$.
}\label{tau}
\end{figure}

In figure \ref{tau}, we represent the resulting 
optical thickness of the accretion disks, 
which are obtained by the method described in \S\ref{thick}, 
as functions of the mass-accretion rate.
As shown in this figure, 
the disks are very optically thick 
in the upper stable branch (burst state), 
$\dot{M}\gsim 10^{-6}M_\odot {\rm yr}^{-1}$.
Hence, the spectral energy distributions (SEDs) 
are though to consist of the superposition
of the blackbody spectra with various temperatures
at the disk surface.
We show them in \S \ref{discussion}.

In contrast, as mentioned above, the POP III disks 
are optically thin
in the regime of $\dot{M}\lsim 10^{-7}M_\odot {\rm yr}^{-1}$.
Such results are unphysical,
since our method described in \S\ref{thick} is available 
only if the disks are optically thick.
The POP I disks are optically thick,
even in the case of $\dot{M}\lsim 10^{-7}M_\odot {\rm yr}^{-1}$.
Cannizzo and Wheeler (\yearcite{Cannizzo84}) reported that POP I disks 
become optically thin for a small mass-accretion rate
in the case of $\alpha\gsim 0.1$, and disks with 
smaller $\alpha$ stay
optically thick down to a far lower mass-accretion rate.
Since $\alpha$ is around $0.03$ in the present work,
our results are consistent with their study.

In the middle unstable branch, where 
$10^{-7}M_\odot {\rm yr}^{-1} \lsim \dot{M}$
$\lsim 10^{-6}M_\odot {\rm yr}^{-1}$ is satisfied,
the optical thickness increases very rapidly with an increase of 
the mass-accretion rate.
Hence, even in the case of POP III disks,
the optical thickness is very large
in the upper stable and middle unstable branches.


It is found that the spiky structure appears around 
$\dot{M}=10^{-8.5}M_\odot {\rm yr}^{-1}$
for $Z/Z_\odot=10^{-4}$. 
The feature is due to the line opacity for heavy elements. 
Such opacity is very sensitive to the temperature as well 
as the density.
Since the matter does not contain the heavy elements for $Z=0$, 
and since dust cooling is dominant over line cooling for 
$Z=Z_\odot$,
the spiky structure appears only in the case of intermediate
metallicity.

\section{Discussion}
\label{discussion}
\subsection{SED of POP III disks}
\label{detection}
We have so far discussed S-shaped equilibrium curves,
which predict the potential possibility of limit-cycle behaviour in POP III
disks via thermal and secular instabilities.
We are now ready to discuss whether the limit-cycle oscillations are
activated in realistic environments surrounding POP III disks, or not.


If we consider  
black holes in halos with 
$M_{\rm halo}\sim 10^8 M_\odot$ at $z\sim 20$, 
the virial temperatures of these halos are 
typically $T\sim 10^4$K, from which the
gas is not evacuated by radiative feedback of 
the progenitor star  \citep{KYSU04}.

The density of the gas surrounding the black holes would be comparable
to, 
or larger than that of the virialized halos \citep{KYSU04}.
Thus, in the first place, 
we assess the mass-accretion rate from the host halo to the
accretion disk by the Bondi accretion rate 
with the given virialized density/temperature of the halo.
\begin{figure}
  \begin{center}
    \FigureFile(90mm,90mm){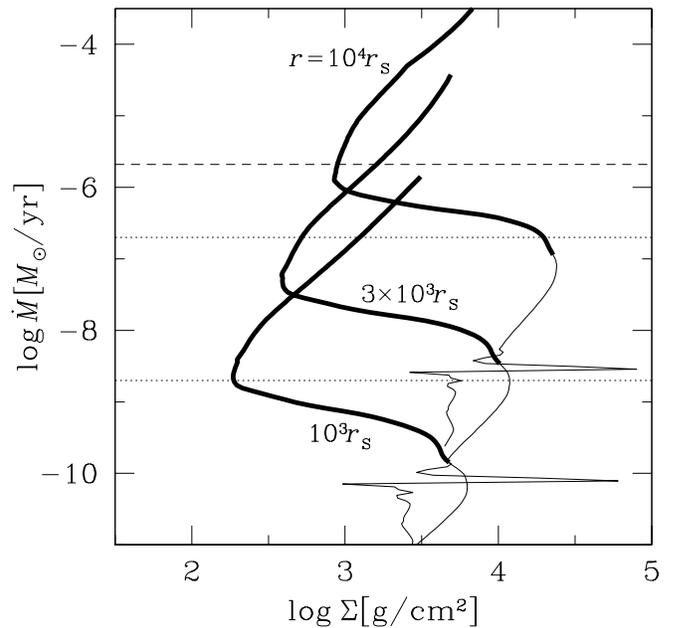}
  \end{center}
  \caption{
 Sequence of equilibrium curves of the accretion disks
 for $r=10^3r_{\rm S}$ ({\it the lower solid lines}),
 $r=3\times 10^3r_{\rm S}$ ({\it the middle solid lines}),
 and $10^4r_{\rm S}$ ({\it the upper solid lines}) 
 in the case of $Z/Z_\odot=10^{-4}$.
 Here, we use the method described in \S\ref{thick}, 
 although the disks are assumed to be optically thick in this method.
 The thick and thin solid lines are equilibrium loci 
 for the disks with $\tau\geq 3$ and $\tau<3$,
 respectively.
 Two dotted lines indicate the Bondi accretion rates
 for $\rho(\infty)=10^{-22}{\rm g \, cm^{-3}}$ ({\it upper})
 and $10^{-24}{\rm g \, cm^{-3}}$ ({\it lower}).
 The critical accretion rate is represented by the dashed line.
 The adopted other parameters are $M=10^3M_\odot$ and
 $c_s(\infty)=10{\rm km \, s^{-1}}$.
  }\label{bondi}
\end{figure}

In figure \ref{bondi},
by using the method described in \S\ref{thick},
we plot the resulting sequence of equilibrium curves 
of disks 
for $r=10^3r_{\rm S}$, $3\times 10^3r_{\rm S}$, 
and $10^4r_{\rm S}$
with $Z/Z_\odot=10^{-4}$
({\it the solid lines}). 
Here, the disks are optically thick, $\tau\geq 3$, 
on the thick solid lines.
In contrast, our method described in \S\ref{thick} 
gives the optically thin solutions 
below the lower turning points
(see the thin solid lines),
although this method is available only 
for optically thick disks.
We also show the Bondi accretion rates for 
$\rho(\infty)=10^{-24} {\rm g \,cm^{-3}}$ 
({\it the lower dotted line})
and $10^{-22} {\rm g\,cm^{-3}}$ 
({\it the upper dotted line}).
We remark that $\rho(\infty)=10^{-24} {\rm g \,cm^{-3}}$ 
is the typical gas density of halos formed 
at $z\sim 10-30$ in standard $\Lambda$ CDM cosmology. 
Here, we also 
assume the sound velocity of $c_s(\infty)=10~{\rm km\,s^{-1}}$,
which corresponds to the halo virial temperature, $10^4$K.
As shown in this figure, 
the S-shaped curve shifts toward upper right 
as the radius increases.
It is obvious that these accretion disks
would exhibit a limit-cycle behaviour
in the regions of $10^{3}r_{\rm S}-10^{4}r_{\rm S}$,
since the Bondi accretion rate corresponds to 
the middle unstable branches at such regions.

We also investigate the possibility to detect the POP III disks.
For this purpose, we try to obtain the disk SEDs 
in the burst state. Since the disks are very optically thick
in this state (see figure \ref{tau}),
we calculate the SEDs by the superposition
of the blackbody spectra with various temperatures at the disk surface.
Here, we consider the limit-cycle oscillation 
induced by disk instability at $r\sim 10^4r_{\rm S}$.
We set the mass-accretion rate 
to be $2.1\times 10^{-7} M_\odot {\rm yr}^{-1}$  
for $r>10^4r_{\rm S}$, 
which is a feasible rate if we assume Bondi accretion
from a small dark halo.
The mass-accretion rates within $10^4r_{\rm S}$ are assumed to be
$\dot{M}=4.5\times 10^{-4} M_\odot {\rm yr}^{-1}$.

We can estimate the temperature profile of the disk surface
using above mass-accretion rates and equation (\ref{Teff}),
except for the innermost regions.
This equation gives the temperature profile of 
$T \propto r^{-3/4}$.
The mass-accretion rate
exceeds the critical rate within $r=10^4r_{\rm S}$
(see figure \ref{Sji}).
Therefore, 
photon trapping becomes nonnegligible 
in the innermost regions of the disk,
$r \lsim (\dot{M}c^2/L_{\rm E})r_{\rm S}$.
In this region, the temperature profile should be modified
as $T \propto r^{-1/2}$.
Such a profile has been reported by a self-similar solution
of the slim disk model \citep{WZ99,WF99}.
\citet{Watarai00} has also shown a similar profile 
by solving the global solution.
\begin{figure}
  \begin{center}
    \FigureFile(90mm,90mm){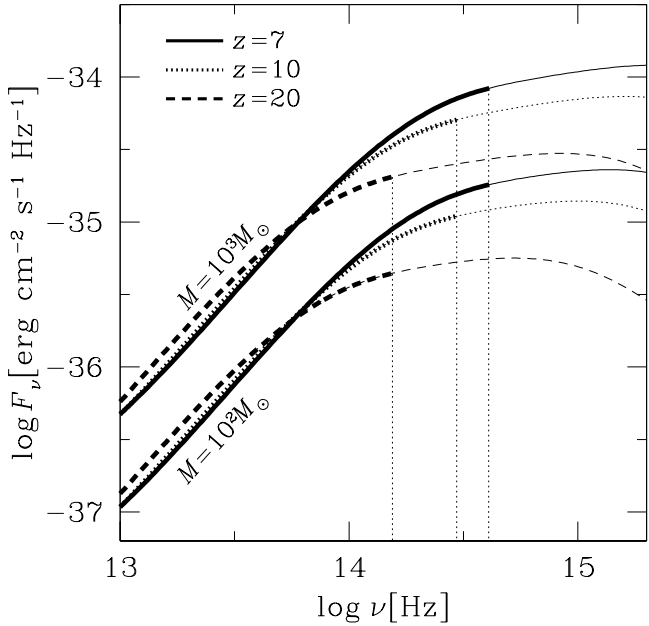}
  \end{center}
  \caption{
 SEDs of POP III black hole accretion disks with $M=10^3M_\odot$ 
 ({\it the three upper lines})
 and $M=10^2M_\odot$ ({\it the three lower lines}). 
 Each line type corresponds to $z=7$ ({\it the solid line}), 
 $10$ ({\it the dotted lines}),
 and $20$ ({\it the dashed lines}).
 The SED break denoted by thin dotted lines 
 at $10^{14}{\rm Hz}\lsim\nu\lsim 10^{15}{\rm Hz}$ represents
 the edge of a Gunn-Peterson trough.
  }\label{sed}
\end{figure}

Figure \ref{sed} represents the SEDs of face-on POP III disks
at three redshifts: $z=7$, $10$ and $20$.
The adopted cosmological parameters are
$\Omega_{\rm M}=0.3$, $\Omega_{\rm V}=0.7$,
and $h_{0.7}=1$.
We also add a break due to the Gunn-Peterson effect for each SED 
at $10^{14}{\rm Hz}\lsim\nu\lsim 10^{15}{\rm Hz}$.
The resulting SED shifts toward lower left
as the redshift increases.

In this figure, we also plot the SEDs of 
POP III disks around black holes with $M=10^2M_\odot$.
We find that the equilibrium curve with $M=10^2M_\odot$
at $r=5\times 10^4r_{\rm S}$
roughly overlaps with the curve for $M=10^3M_\odot$ at $r=10^4r_{\rm S}$.
Thus, we consider the limit-cycle oscillation occurs at 
$r\sim 5\times 10^4r_{\rm S}$.
The mass-accretion rates at $r\leq 5\times 10^4r_{\rm S}$
and at $r>5\times 10^4r_{\rm S}$
are set to be 
$\dot{M}=4.5\times 10^{-4} M_\odot {\rm yr}^{-1}$
and $\dot{M}=10^{-7}M_\odot {\rm yr}^{-1}$,
respectively.
We find that the flux densities in the case of 
$M=10^2M_\odot$ 
are about $20-30\%$ of those in the case of $M=10^3M_\odot$
at $10^{13}{\rm Hz}\lsim \nu \lsim 10^{15}{\rm Hz}$.

Those flux densities are quite low for available facilities at
present; however, 10 hr integration by OWL telescope manages to reach
the flux for $10^3M_\odot$ around $1\mu m$, 
if they could achieve the same 
limiting flux at $1\mu m$ as V-band
\footnote{According to the project
URL(http://www.eso.org/projects/owl/\\index\_2.html), OWL will have
limiting magnitude of ${\rm V}\sim 38$.}
.



\subsection{Number of Sources}
Here, we evaluate the expected number density of POP III BHs. 
The number density of POP III BHs could be evaluated by the
star-formation rate of POP III stars. Theoretical studies  \citep{Sokasian04}
predict that the comoving star-formation density is roughly $\sim 10^{-3}M_\odot~{\rm yr}^{-1}~{\rm Mpc}^{-3}$ at $z\gsim 20$, although
there are still large uncertainties due to various feedback effects 
(e.g, \authorcite{SU06} \yearcite{SU06}).
If we assume all of the POP III stars turned out to be POP III BHs, 
the number density of POP III BHs is roughly evaluated as follows:
\begin{eqnarray}
 n_{\rm P3BH}=& & 1.0 \times 10^2 
  \left(\frac{SFR}{10^{-3}M_\odot{\rm yr}^{-1}
 {\rm Mpc}^{-3}}\right) \nonumber\\
  & & \times \left(\frac{t}{10^8 {\rm yr}}\right)
  \left(\frac{M}{10^3M_\odot}\right)^{-1}{h}_{0.7}^{3}{\rm Mpc}^{-3}.
  \label{eq:rhoSFR}
\end{eqnarray}
Here, $t$ denotes the duration of POP III star formation.

On the other hand, the comoving volume, which corresponds to the observed
field of view with $\theta\times\theta$ at $z \sim z+\Delta z$, is 
\begin{eqnarray}
V=& &1.6\times 10^4\left(\frac{\theta}{10'}\right)^2\left(\frac{\Delta
      z}{0.1}\right)\nonumber\\
 & &\times \left[ \frac{\phi\left(z\right)}{1} \right]
      \left(\frac{1+z}{11}\right)^{-1}{h}_{0.7}^{-3}{\rm Mpc}^3.
\label{eq:vol}
\end{eqnarray}
Here, $\phi(z)$ is a function that equals to almost unity for $z\gsim 10$.
Finally, we obtain the observed number of sources by multiplying 
$n_{\rm P3BH}$, $V$ and the efficiency $\epsilon$ 
(see last paragraph in \S\ref{sequence}):
\begin{eqnarray}
N_{\rm obs}=& & 8.0\times 10^3
\left(\frac{\theta}{10'}\right)^2
\left(\frac{\Delta z}{0.1}\right)
\left(\frac{\epsilon}{5\times 10^{-3}}\right) \nonumber\\
& & \times
\left[ \frac{\phi\left(z\right)}{1}\right]
\left(\frac{1+z}{11}\right)^{-1}.
\end{eqnarray}
Although these are optimistic assessments, it seems possible to
find $10^3 M_\odot$ BHs at $z=10$ if we have telescopes that cover a
$10'\times 10'$ field of view with a sufficient limiting magnitude.

As supplementary evidence, we also evaluate the number density
of POP III BHs based on the local SMBH density.
Yu and Tremaine (\yearcite{YT02}) assessed
the density of massive black holes in the present universe using a
simple correlation between the black hole mass and the velocity dispersion 
of the bulge stars \citep{Gebhardt00,FM00}. 
If we consider most of the
POP III BHs are taken in by massive black holes through merging,
present-day massive black hole density gives the upper limit of the
POP III BH density. In fact, POP III BHs are formed at high-density 
peaks of a random Gaussian density fluctuation of the universe,
which tend to be located within density fluctuations with larger
scales, such as galactic scales. Such intermediate-mass black holes in a
galaxy settle onto the galactic centre, since they lose their orbital
angular momentum through dynamical friction and 
three-body interaction processes \citep{Ebi01}. 

According to Yu and Tremaine (\yearcite{YT02}),
the local massive black hole density is given by 
\begin{equation}
\rho_{\rm MBH}=2.9\times 10^5 {h}_{0.7}^2 M_\odot {\rm Mpc}^{-3}. 
\label{eq:rhoMBH}
\end{equation}
Thus, the upper limit of the comoving POP III BHs number density at high
redshift is 
\begin{equation}
n_{\rm P3BH}=2.9\times 10^2 \left(\frac{M}{10^3M_\odot}\right)^{-1}
h_{0.7}^2 {\rm Mpc}^{-3}.
\label{eq:rhoMBH}
\end{equation}
It is suggestive that this predicted number density is only a few-times
larger than the value given in equation (\ref{eq:rhoSFR}).


\section{Conclusions}
By solving the vertical structure of the optically-thick 
accretion disks,
including the convective energy transport,
and by employing the one-zone model for optically thin
isothermal disks,
we investigate the stability of the disks 
in primordial environments.

POP III disks ($Z\lsim 10^{-4}Z_\odot$)
are thermally and secularly unstable,
and exhibit limit-cycle oscillations, like the POP I disks
($Z\sim Z_\odot$).
The outbursts of POP III disks are stronger than 
those of POP I disks.
The maximal luminosity in the burst state 
is expected to be an order of magnitude larger 
in POP III disks than in POP I disks.


The expected flux densities of POP III disks 
surrounding black holes of $M=10^3M_\odot$
are 
$\sim 10^{-34} {\rm erg~s^{-1}~cm^{-2}~Hz^{-1}}$ 
around $1\rm{\mu m}$ in the case of $z\sim 10$, 
whereas the expected number of sources is
$O(10^3)$ per $10'\times 10'$ field of
view. Those disks are within reach of future 
facilities, such as OWL.


\bigskip
\bigskip

Numerical calculations were carried out at Rikkyo University.
This work was supported in part 
by a special postdoctoral researchers program in RIKEN (KO),
and by Ministry of Education, Culture,
Sports, Science, and Technology (MEXT) 
Young Scientists (B) 17740111(KO) and 17740110 (HS).

\end{document}